\documentclass[useAMS]{mn2e}
\usepackage{graphics,graphicx,array,amssymb}

\title[Consistent Baade-Wesselink distances]
{Consistent distances from Baade-Wesselink analyses of Cepheids 
and RR~Lyraes}
\author[G. Kov\'acs]{G\'eza Kov\'acs$^{1}$\thanks{E-mail: kovacs@konkoly.hu}, 
\\
$^1$Konkoly Observatory, PO Box 67, 1525 Budapest, Hungary\\
}
\begin{document}

\date{Received day month 2003 / Accepted day month 2003}

\pagerange{\pageref{firstpage}--\pageref{lastpage}} \pubyear{200?}

\maketitle

\label{firstpage}

\begin{abstract}
By using the same algorithm in the Baade-Wesselink analyses of 
Galactic RR~Lyrae and Cepheid variables, it is shown that, within 
$0.03$~mag $1\sigma$ statistical error, they yield the same distance 
modulus for the Large Magellanic Cloud. By fixing the zero point of 
the color--temperature calibration to those of the current infrared 
flux methods and using updated period--luminosity--color relations, 
we get an average value of $18.55$ for the true distance modulus of 
the LMC. 
\end{abstract}


%%%%%%%%%%%%%%%%%%%%%
%  SECTION 1
%%%%%%%%%%%%%%%%%%%%%

\section{Introduction}
A large number of papers have been published lately on the determination 
of the distance to the Large Magellanic Cloud (LMC). This increased 
interest is attributed to the extreme values published after the 
analyses of the {\sc hipparcos} data on Cepheids (Feast \& Catchpole 
1997) and to the application of the red clump stars of the 
LMC through the same type of Galactic stars calibrated by 
{\sc hipparcos} (Stanek et al. 2000). Along this line, there seems to 
exist a general belief that Cepheids yield usually longer distances 
(distance moduli of about $18.5$~mag and above), whereas RR~Lyrae 
stars give shorter ones ($\approx 18.3$~mag). This latter belief is 
based on statistical parallaxes (Popowski \& Gould 1998), Baade-Wesselink 
(BW) analyses (see Clementini et al. 1995, and references therein, 
hereafter C95) and the application of the {\sc hipparcos} data 
(Fernley et al. 1998; see however Groenewegen \& Salaris 1999). 
It is important to note that distances derived from the statistical 
parallax method employ rather small sample of stars ($\approx 200$ 
objects), whereas the {\sc hipparcos} parallaxes for RR~Lyrae stars 
are very inaccurate. Actually, more precise HST parallaxes yield 
distances in good agreement with the "canonical" LMC value of 
$18.5$~mag derived from both prototype variables RR~Lyr and $\delta$~Cep 
(Benedict et al. 2002a,b). It is also important to mention 
that all results based on the pulsation analyses of double-mode 
variables (RR~Lyrae, Cepheids, both fundamental \& first overtone 
and first overtone \& second overtone ones) give basically the 
canonical value for the LMC distance (see Kov\'acs 2002 and 
references therein).

In the light of the above, it is important to address the question of 
consistency between the BW analyses of Cepheids and RR~Lyrae stars.  
Previous works have been performed by different groups of researchers, 
often using different BW parameters (e.g., $p$ factors, the ratio of the 
observed and surface radial velocities) and effective temperature relations 
(surface brightness for Cepheids -- e.g., Laney \& Stobie 1994, stellar 
atmosphere models for RR~Lyrae stars -- e.g., Liu \& Janes 1990). In 
addition, there is a role of the somewhat delicate nature of the numerical 
method employed in the actual implementation of the BW analysis (e.g., 
computation of the surface acceleration). Therefore, we decided to visit 
the problem by using the same BW algorithm for both groups of stars. In 
order to compute the LMC distance modulus, the derived absolute magnitudes 
will be used in conjunction with currently published Period-Luminosity-Color 
(PLC) relations (Udalski et al. 1999; Kov\'acs \& Walker 2001, hereafter 
U99 and KW01, respectively).

%%%%%%%%%%%%%%%%%%%%%
%  SECTION 2
%%%%%%%%%%%%%%%%%%%%%

\section{Implementation of the BW method}
Here we basically follow the standard methodology of the BW analysis 
as usually implemented in RR~Lyrae studies (e.g., Liu \& Janes 1990). 
In brief, we use the observed $V$ light, $V-K$ color and $V_{\rm rad}$ 
radial velocity curves to compute the luminosity ($L$), effective 
temperature ($T_{\rm eff}$) and radius ($R$) variations. For any 
assumed value of the equilibrium test radius $R_0$ we compute the 
following quantities: 
\begin{itemize}
\item
{\it Temporal radius:} 

\begin{itemize}
\item
$R(t)=R_0+\Delta R(t)$,
\item
$\Delta R(t)=-p\int_0^t[V_{\rm rad}(\tau)-\langle V_{\rm rad} \rangle]d\tau$, 
\item
$\langle V_{\rm rad} \rangle = P_0^{-1}\int_0^{P_0} V_{\rm rad}(\tau)d\tau$,
\end{itemize}

\noindent
where $P_0$ denotes the period, $p$ is the velocity projection factor, 
that we assume to be constant and set equal to $1.35$. (We note that 
Gieren, Fouqu\'e \& G\'omez 1997 used a period-dependent $p$ factor. 
Their values span the range of 1.35--1.37 in the period regime of our 
sample. The effect of using lower $p$ is minute, because an increase 
of $0.02$ in $p$ yields an absolute magnitude that is brighter only 
by $0.02$~mag.)

\item
{\it Temporal $T_{\rm eff}$:}

\begin{itemize}
\item
$\log T_{\rm eff} = f[(V-K)_0, {\rm [Fe/H]}, \log g]$,
\end{itemize}

\noindent
where $f[...]$ denotes the interpolation polynomial of the stellar 
atmosphere models of Castelli, Gratton \& Kurucz (1997) with the 
zero point determined by the scale of Blackwell \& Lynas-Gray, (1994) 
(hereafter BLG94). The models have been computed without convective 
overshooting. The microturbulent velocity has been fixed at 
2~km\thinspace s$^{-1}$. The distribution of heavy elements follows 
that of the Sun. In order to find the temperature zero point offset 
between the model values of Castelli et al. (1997) and the ones 
given by the infrared flux method of BLG94, the compilation of the 
observational data on non-variable stars by C95 is used. We get 
$\log T_{\rm eff}$(model)$-\log T_{\rm eff}$(BLG94)$=0.0036$.
 
It is interesting to compare this temperature scale with the one 
implied by the surface brightness (SB) method of Fouqu\'e \& Gieren (1997). 
Rewriting their formula given for the apparent angular diameter, 
we get $\log T_{\rm eff}=-0.131(V-K)_0-0.1BC+3.9474$. By using the 
above mentioned compilation of C95, we get an average difference 
of $\log T_{\rm eff}$(SB)$-\log T_{\rm eff}$(model\& BLG94)$=-0.0035$. 
By performing the same test on the BW samples (see Table~1), the 
differences are positive, but less than $\approx 0.002$ for Cepheids. 
For RR~Lyrae stars, the differences are even smaller, with changing 
signs. The larger negative difference on the sample of C95 is due 
in part to the larger weight of the higher temperature stars. In 
conclusion, we may state that the present temperature scale in the 
temperature range of interest is very similar to the ones used in 
earlier BW studies of Cepheids.       

For RR~Lyrae stars, the iron abundance [Fe/H] is computed from the 
Fourier decompositions of the $V$ light curves as given by Jurcsik 
\& Kov\'acs (1996). For Cepheids, we set [Fe/H]$=$0, which seems 
to be a reasonable first approximation (see Luck et al. 2003 for 
the metallicity distribution of Cepheids). The temporal gravity 
$g$ is computed by $g=GM/R^2-p\thinspace dV_{\rm rad}/dt$, where 
$G$ is the gravitational constant, $M$ is the stellar mass, which 
is fixed at $0.7M_{\sun}$ and $5.0M_{\sun}$ for RR~Lyrae stars and 
Cepheids, respectively. Since the contribution of $\log g$ to 
$\log T_{\rm eff}$ is relatively small when the $V-K$ color index 
is used, the exact value of $M$ is not crucial. In principle, we 
can compute a consistent value for $M$ by using the pulsation 
equation (e.g., van Albada \& Baker 1973). However, both the weak 
dependence of $\log T_{\rm eff}$ on $\log g$, and the difference 
between the equilibrium and static stellar parameters (see Bono, 
Caputo \& Stellingwerf 1995) make this procedure less straightforward 
and therefore, is left for the task of future studies. The dereddened 
color index $(V-K)_0$ is computed from the observed one by using 
$E_{B-V}$ reddening values either as given in the Galactic Cepheid 
Database of the David Dunlap 
Observatory\footnote{http://ddo.astro.utoronto.ca/cepheids.html} 
or by the corresponding formula of KW01 for RR~Lyrae stars. 
Because of the large reddenings of the Galactic Cepheids, we use 
the formulae for selective absorption ratios as given by the BW 
studies of Cepheids (see, e.g., Gieren, Fouqu\'e \& G\'omez 1998): 

\begin{itemize}
\item
$R_V = 3.07 + 0.28(B-V)_0 + 0.04E_{B-V}$,
\item
$R_I = 1.82 + 0.205(B-V)_0 + 0.022E_{B-V}$,
\item
$R_K = 0.279$,
\item
$A_V=R_VE_{B-V}$, $A_I=R_IE_{B-V}$, $A_K=R_KE_{B-V}$,
\end{itemize}

\noindent
where the $I$ and $K$ colors are in the Kron-Cousins and standard 
Johnson systems, respectively. For compatibility, these reddening 
relations are used also for RR~Lyrae stars. 

\item
{\it Temporal absolute $V$ magnitude:}

\begin{itemize}
\item
$M_V = -2.5\log L + M_{\rm bol}^{\sun} - 
BC(\log T_{\rm eff},{\rm [Fe/H]},\log g)$,
\item
$\log L = 4\log T_{\rm eff}/T_{\rm eff}^{\sun} + 2\log R/R_{\sun}$,
\end{itemize}

\noindent 
where $R_{\sun}$ and  $T_{\rm eff}^{\sun}$ are the solar values. 
The latter one and the solar bolometric magnitude $M_{\rm bol}^{\sun}$ 
are set equal to $5778$~K and $4.75$~mag, respectively. The temporal 
bolometric correction $BC$ is computed from stellar atmosphere models 
in the same way as $\log T_{\rm eff}$ above. The zero point of $BC$ is 
set to a value which yields $BC_{\sun}=-0.07$.
 
\item
{\it Point-by-point and average distance moduli:}

\begin{itemize}
\item
$DM(i) = V(i) - R_V E_{B-V}(i) - M_V(i)$,
\item
$\langle DM\rangle = N^{-1}\sum_{i=1}^N DM(i)$,
\item
$\sigma_{DM}^2 = (N-1)^{-1}\sum_{i=1}^N [DM(i)-\langle DM \rangle]^2$,
\end{itemize}

\noindent
where $N$ is the number of data points on the light curve. 

\end{itemize}

The best fit is obtained at the test radius $R_0$ which yields the 
smallest $\sigma_{DM}$. Usually the fits are fairly accurate, with 
$\sigma_{DM}=0.02$--$0.01$. It is also worth mentioning the following 
technical details:

\begin{itemize}
\item
When straightforward Fourier fitting does not work, smooth light and 
velocity curves are obtained through Fourier decompositions of the 
high order polynomials fitted by least squares to the observed data 
points of the folded light curves.
\item
Tables of stellar atmosphere models are employed through quadratic 
interpolation.
\item
All pulsation phases are used, including also the possibly shock-disturbed 
quick expansion phase.
\item
In spite of the careful treatment of the numerical derivation of the 
radial velocity curve, the total gravitational acceleration occasionally 
reaches values which lie outside the tabulated values. In these brief 
moments of pulsation the closest extreme values as given in the tables 
are used. The same method is employed for filtering out the non-tabulated 
values of $T_{\rm eff}$. We note that these out-of-range errors occur   
almost always out of the minimum of $\sigma_{DM}$, where the assumed 
unphysical stellar radius further enhances the numerical problems  
related to $\log g$.  
\end{itemize}

%%%%%%%%%%%%%%%%%%%%%
%  SECTION 3
%%%%%%%%%%%%%%%%%%%%%

\section{PLC relations}

Once the optimum equilibrium stellar parameters are computed by 
the above BW technique, we can evaluate the zero points of the 
corresponding PLC relations. Both for Cepheids and for RR~Lyrae 
stars the PLC relations have been derived from stellar populations 
different from the Galactic BW samples. In the case of RR~Lyrae 
stars there is a reasonable large overlap between the metallicities 
of the present BW sample and the cluster variables used by KW01 to 
derive the empirical formula. Therefore, we may assume that the 
formula may also be valid for most of the field variables used 
in the BW analysis. There is also a dispute on the possible metal 
dependence of the zero point of the Cepheid PLC relation. However, 
we think that there are more theoretical and empirical results 
which indicate the non-existence of such a dependence (Alibert et al. 
1999; Bono et al. 2002; Turner \& Burke 2002).    

In the case of Cepheids we need to evaluate $C_{VI}$ in the following 
equation
%
%
%>>>>>>>>>>>
%  EQ. 1
%>>>>>>>>>>>
%
\begin{eqnarray}
\langle I_c \rangle - 1.55(\langle V \rangle - \langle I_c \rangle ) 
= -3.300\log P_0 + C_{VI} \hskip 3mm ,
\end{eqnarray}
where $I_c$ denotes the Kron-Cousins colors, $\langle ...\rangle$ means 
intensity-averaged magnitudes. The above formula has been derived by 
U99 for LMC with $C_{VI}=15.868$. Since the left-hand-side of Eq.~(1) is 
reddening-free, the difference of $C_{VI}$(LMC) and the average of this 
constant for the BW stars will yield the true distance modulus for the 
LMC.

The PLC relation for the fundamental mode RR~Lyrae stars is given by 
KW01
%
%
%>>>>>>>>>>>
%  EQ. 2
%>>>>>>>>>>>
%
\begin{eqnarray}
 \overline V - 3.1(\overline B - \overline V) = 
-2.467\log P_0 + C_{BV} \hskip 3mm ,
\end{eqnarray}
where the bars denote simple magnitude averages. The constant for LMC 
could be evaluated from the current $V$, $B$ observations of fundamental 
mode RR~Lyrae stars in the LMC by Clementini et al. (2003, hereafter C03). 
Because they give intensity-averaged magnitudes, to retain compatibility 
with Eq.~(2), we transform these values to simple magnitude averages 
with the aid of the formulae of Kov\'acs (2002). The average of the 
$A_1$ Fourier amplitude is set equal to $0.30$, in accordance with 
the value obtained from the analysis of the RRab stars in the 
corresponding {\sc macho} fields (Alcock et al. 2003). In this way we get  
$\overline V = 0.032 + \langle V \rangle$ and
$\overline B = 0.058 + \langle B \rangle$. With the observed average 
$P_0$, $\langle V \rangle$ and $\langle B \rangle$ of $0\fd58$, $19.366$ 
and $19.744$, respectively, we get $C_{BV}=17.56$. However, the average magnitudes computed by C03 for their two fields $A$ \& $B$ differ by as 
much as $0.08$~mag in $V$, and $0.13$~mag in $B$. They attribute this 
difference to the effect of differential extinction. We checked this 
difference by computing the averages in {\sc macho} fields \#6 \& 
\#13, overlapping with fields $A$ \& $B$ of C03. With a $4\sigma$ 
clipping we got 
$\langle V \rangle = 19.311\pm 0.012$ (field \#6, 317 stars) and 
$\langle V \rangle = 19.313\pm 0.010$ (field \#13, 262 stars). 
With these small standard deviations of the means, we can state 
that a difference in the order of $0.1$~mag is not probable between 
the two fields.  

In order to get another estimate on $C_{BV}$ for the LMC, we use the 
data of Soszy\'nski et al. (2002), which cover a large sample of RR~Lyrae 
stars of the Small Magellanic Cloud (SMC) from the {\sc ogle} database. 
By proceeding as above, from their average magnitudes and period of 
$\langle V \rangle = 19.74$, $\langle B \rangle = 20.10$, 
$\langle V_0 \rangle = 19.45$, $\langle B_0 \rangle = 19.73$ and 
$\langle P_0 \rangle = 0.59$, we get $C_{BV}(SMC)=18.01$ and $17.97$, 
depending on whether reddened or dereddened magnitudes are used. 
Although Eqs.~(1) and (2) should, in principle be reddening-free, 
the above difference shows that the field-by-field reddening corrections 
applied by Soszy\'nski et al. (2002) may not be completely consistent 
with our standard extinction ratio. Since the method of their differential 
reddening correction has already proved to be successful (at least in 
statistical sense, see U99), we adopt the value obtained by the use of 
their dereddened magnitudes. In computing the value of $C_{BV}$ for 
the LMC, we have to subtract the relative distance modulus between 
the LMC and SMC. From the PLC relations derived for Cepheids in the 
two clouds, U99 got a difference of $0.51$~mag, which, with the 
above preliminaries, yields our finally adopted value of $C_{BV}=17.46$ 
for the LMC.

%%%%%%%%%%%%%%%%%%%%%
%  SECTION 4
%%%%%%%%%%%%%%%%%%%%%

\section{LMC distances from BW absolute magnitudes and PLC relations}

The necessary data on RR~Lyrae stars have been gathered from the 
literature during our earlier investigations (see KW01 and also 
C95, for a full list of variables subjected to BW analyses in earlier 
studies). For Cepheids we turned to the database 
of the McMaster University\footnote{http://dogwood.physics.mcmaster.ca/Cepheid//HomePage.html}
and also to the radial velocity database of the Moscow University 
(see Gorynya et al. 1998). We utilized only the best data available 
on fundamental mode variables. When it was necessary, the Cepheid 
$I$ light curves published in the Johnson system were transformed to 
the Kron-Cousins system with the aid of the formula given by Caldwell 
\& Coulson (1987). Occasionally, periods have been improved by using 
the program package {\sc mufran} (see Koll\'ath 1990). The variables 
used in the present study together with their derived physical parameters 
are given in Table~1. (Effective temperatures have been computed from 
the cycle-averaged $V-K$ and $\log g$ values.) Three RR~Lyrae stars 
(W~Crt, SS~Leo and AV~Peg) have been excluded from the above sample 
in the subsequent analysis, because of their discrepant positions in 
the PLC relation. For similar reasons, we also excluded X~Lac from the 
Cepheid sample. 

%
%%%%%%%%%%%%
%  TABLE 1
%%%%%%%%%%%% 
%

\begin{table}
\caption[]{Derived physical parameters of the BW variables}
\label{typical}
\begin{center}
\begin{tabular}{lrrrrr}
\hline
\multicolumn{6}{c}{RR~Lyraes} \\
\hline
\multicolumn{1}{l}{Name} &
\multicolumn{1}{r}{$\log P$} &
\multicolumn{1}{c}{$T_{\rm eff}$} &
\multicolumn{1}{r}{$R/R_{\sun}$} &
\multicolumn{1}{r}{$M_V$} &
\multicolumn{1}{r}{$C_{BV}$} \\
\hline
SW And    & $-0.354$ & 6702 & $  4.47$ & $ 0.81$ & $-1.09$ \\ 
WY Ant    & $-0.241$ & 6296 & $  6.10$ & $ 0.52$ & $-1.17$ \\ 
X Ari     & $-0.186$ & 6111 & $  6.85$ & $ 0.43$ & $-1.15$ \\ 
RR Cet    & $-0.257$ & 6420 & $  5.22$ & $ 0.75$ & $-0.98$ \\ 
UU Cet    & $-0.217$ & 6258 & $  6.25$ & $ 0.48$ & $-1.22$ \\ 
W Crt     & $-0.385$ & 6788 & $  4.90$ & $ 0.56$ & $-1.31$ \\ 
DX Del    & $-0.325$ & 6691 & $  4.62$ & $ 0.75$ & $-1.13$ \\ 
SU Dra    & $-0.180$ & 6293 & $  5.85$ & $ 0.62$ & $-0.94$ \\ 
SW Dra    & $-0.244$ & 6393 & $  5.43$ & $ 0.68$ & $-1.02$ \\ 
RX Eri    & $-0.231$ & 6383 & $  5.40$ & $ 0.70$ & $-0.98$ \\ 
RR Gem    & $-0.401$ & 6721 & $  4.60$ & $ 0.74$ & $-1.19$ \\ 
TW Her    & $-0.398$ & 6732 & $  4.55$ & $ 0.76$ & $-1.16$ \\ 
RR Leo    & $-0.344$ & 6458 & $  4.85$ & $ 0.87$ & $-0.93$ \\ 
SS Leo    & $-0.203$ & 6431 & $  6.80$ & $ 0.16$ & $-1.41$ \\ 
TT Lyn    & $-0.224$ & 6283 & $  5.72$ & $ 0.66$ & $-1.04$ \\ 
V445 Oph  & $-0.401$ & 6792 & $  4.03$ & $ 0.96$ & $-1.02$ \\ 
AV Peg    & $-0.409$ & 6836 & $  3.62$ & $ 1.18$ & $-0.78$ \\ 
AR Per    & $-0.371$ & 6833 & $  4.45$ & $ 0.72$ & $-1.22$ \\ 
BB Pup    & $-0.318$ & 6712 & $  4.40$ & $ 0.87$ & $-0.95$ \\ 
W Tuc     & $-0.192$ & 6316 & $  6.75$ & $ 0.27$ & $-1.26$ \\ 
TU UMa    & $-0.254$ & 6369 & $  5.75$ & $ 0.57$ & $-1.14$ \\ 
UU Vir    & $-0.323$ & 6543 & $  4.95$ & $ 0.75$ & $-1.06$ \\ 
\hline
\multicolumn{6}{c}{Cepheids} \\
\hline
\multicolumn{1}{l}{Name} &
\multicolumn{1}{r}{$\log P$} &
\multicolumn{1}{c}{$T_{\rm eff}$} &
\multicolumn{1}{r}{$R/R_{\sun}$} &
\multicolumn{1}{r}{$M_V$} &
\multicolumn{1}{r}{$C_{VI}$} \\
\hline
$\eta$~Aql   & $ 0.856$ & 5675 & $ 60.8$ & $-4.04$ & $-2.97$ \\ 
TT Aql    & $ 1.138$ & 5589 & $ 76.8$ & $-4.46$ & $-2.61$ \\ 
VY Car    & $ 1.277$ & 5257 & $112.8$ & $-4.95$ & $-2.85$ \\ 
WZ Car    & $ 1.362$ & 5208 & $107.6$ & $-4.78$ & $-2.38$ \\ 
V Cen     & $ 0.740$ & 5684 & $ 42.8$ & $-3.28$ & $-2.60$ \\ 
VW Cen    & $ 1.177$ & 4998 & $ 89.2$ & $-4.12$ & $-2.50$ \\ 
XX Cen    & $ 1.040$ & 5483 & $ 64.0$ & $-3.96$ & $-2.48$ \\ 
$\delta$~Cep & $ 0.730$ & 5765 & $ 44.0$ & $-3.42$ & $-2.71$ \\ 
X Cyg     & $ 1.214$ & 5291 & $100.0$ & $-4.72$ & $-2.77$ \\ 
X Lac     & $ 0.736$ & 5815 & $ 55.6$ & $-3.98$ & $-3.18$ \\ 
CV Mon    & $ 0.731$ & 5689 & $ 40.0$ & $-3.14$ & $-2.59$ \\ 
T Mon     & $ 1.432$ & 5096 & $145.6$ & $-5.31$ & $-2.83$ \\ 
UU Mus    & $ 1.066$ & 5512 & $ 72.0$ & $-4.24$ & $-2.65$ \\ 
U Nor     & $ 1.102$ & 5696 & $ 80.4$ & $-4.67$ & $-2.89$ \\ 
BF Oph    & $ 0.609$ & 5903 & $ 36.8$ & $-3.16$ & $-2.77$ \\ 
BN Pup    & $ 1.136$ & 5475 & $ 74.0$ & $-4.26$ & $-2.53$ \\ 
RY Sco    & $ 1.308$ & 5312 & $ 98.8$ & $-4.72$ & $-2.52$ \\ 
BB Sgr    & $ 0.822$ & 5645 & $ 50.0$ & $-3.58$ & $-2.70$ \\ 
U Sgr     & $ 0.829$ & 5741 & $ 51.6$ & $-3.75$ & $-2.82$ \\ 
WZ Sgr    & $ 1.339$ & 5064 & $125.6$ & $-4.94$ & $-2.72$ \\ 
RY Vel    & $ 1.449$ & 5363 & $132.0$ & $-5.41$ & $-2.66$ \\ 
RZ Vel    & $ 1.310$ & 5210 & $122.8$ & $-5.07$ & $-2.88$ \\ 
SW Vel    & $ 1.370$ & 5231 & $114.0$ & $-4.93$ & $-2.53$ \\ 
T Vul     & $ 0.647$ & 5986 & $ 38.4$ & $-3.32$ & $-2.67$ \\ 
U Vul     & $ 0.903$ & 5964 & $ 55.2$ & $-4.10$ & $-2.90$ \\ 
\hline
\end{tabular}
\end{center}
\end{table}

The corresponding PLC relations are shown in Figs.~1 and 2. The 
constants in Eqs.~(1) and (2) have been derived from the BW absolute 
magnitudes and reddening-free colors. The straight lines represent 
these calibrated relations. It is seen that the empirical slopes of 
the PLC relations fit perfectly the BW data (please note the highly 
different scales of the two figures). This strong correlation is 
very comforting, since two, completely separate methods and datasets 
are compared (BW results, based on Galactic variables on one hand, 
and empirical results, based on LMC and globular cluster variables 
on the other). The standard deviations of the BW points around the 
straight lines are $0.104$ and $0.157$ for RR~Lyraes and Cepheids, 
respectively. We recall that the corresponding empirical values 
are $0.041$ and $0.058$ (see KW01 and U99). The much smaller values 
of these empirical standard deviations show that the BW luminosities 
still suffer from large random errors as it follows from the 
delicate nature of the BW method. 

Computation of the LMC distance modulus is straightforward from 
the above results. It is obtained simply by the subtraction of the 
corresponding constants in the BW and empirical relations. By using 
the results of Sect.~3, from the RR~Lyrae sample we get 
$m-M=17.46+1.09=18.55$, whereas from the Cepheid sample we obtain 
$m-M=15.87+2.69=18.56$. Although these true distance moduli are 
magically close to each other, we have to keep in mind that they both 
have a statistical errors of $\approx 0.03$~mag. Nevertheless, it is 
clear that the often cited $0.2$--$0.3$~mag difference between the 
BW distances derived from the two groups of variables no longer exists. 
They both yield the same distance within a relatively small statistical 
uncertainty. 

%
%%%%%%%%%%%%
%  FIGURE 1
%%%%%%%%%%%% 
%
\begin{figure}
\centering
\includegraphics[angle=0,scale=0.37]{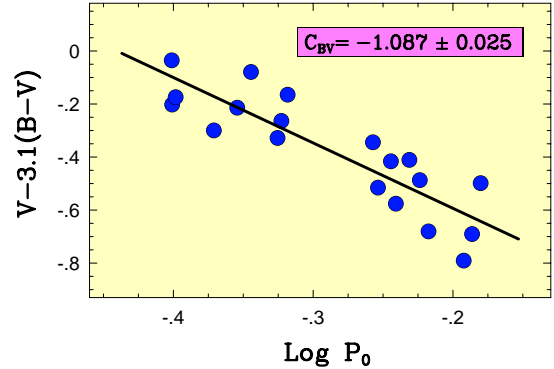}
\caption{PLC relation for the 19 fundamental mode RR~Lyrae stars 
listed in Table~1. The straight line corresponds to Eq.~(2) with 
the adjusted constant given in the upper right box.}
\end{figure}
%

%
%%%%%%%%%%%%
%  FIGURE 2
%%%%%%%%%%%% 
%
\begin{figure}
\centering
\includegraphics[angle=0,scale=0.37]{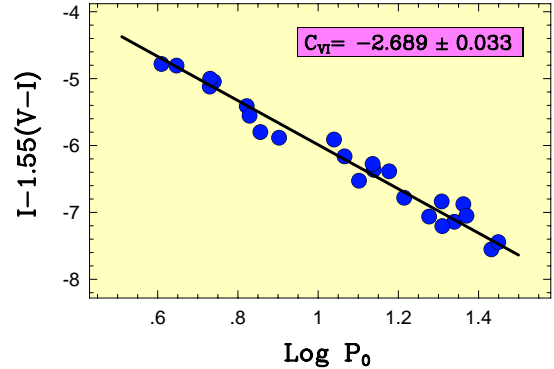}
\caption{PLC relation for the 24 fundamental mode $\delta$~Cephei  
stars listed in Table~1. The straight line corresponds to Eq.~(1) with 
the adjusted constant given in the upper right box.}
\end{figure}

It is interesting to compare the absolute magnitudes obtained 
in the present study with those derived earlier in other BW works. 
With the 18 Cepheids common with the sample of Gieren at al. (1998), 
we get and average difference (ours-theirs) of $-0.003$~mag, with a 
standard deviation $\sigma$ of $0.160$~mag. A similar comparison with 
the absolute magnitudes given for $p=1.35$ in Table~21 of C95, yields 
$-0.162$ with $\sigma = 0.094$. This considerable difference results 
in a proper shift in the RR~Lyrae distance modulus and leads to a 
result which is consistent with the one derived from Cepheids. The 
fact that Gieren et al. (1998) derives a $0.09$~mag lower distance 
for the LMC, in spite of the same average luminosity, is suspected 
to be caused by the difference between the PLC we use and the PL 
they employ to compute the LMC distance. 

We find also a good agreement if we compare our radii with those 
of Gieren et al. (1998). For the 18 common variables we get an average 
difference (ours-theirs) of $-0.4$~$R_{\sun}$ with 
$\sigma=6.3$~$R_{\sun}$. In terms of the relative radii these 
figures become $-0.04$\% and $6.9$\%, respectively.

We can also make a comparison with the results of long-baseline 
interferometric measurements for Cepheids. For the two variables 
$\delta$~Cep and $\eta$~Aql common with the samples of Nordgren et al. 
(2002) and Lane, Greech-Eakman \& Nordgren (2002), we find a rather 
close agreement between the radii. By applying the necessary 
corrections due to the different $p$ factors, from the 
interferometric/radial velocity results we obtain the radii of 
$42\times 1.35/1.31=43R_{\sun}$ and $62\times 1.35/1.43=58R_{\sun}$ for 
$\delta$~Cep and $\eta$~Aql, respectively. These values are very 
close to our corresponding figures (see Table~1).  

It is important to note that although all direct estimates based on 
double-mode variables yield LMC distance modulus in agreement with 
the present result (Kov\'acs 2002), the indirect distances derived 
from globular cluster double-mode RR~Lyrae (RRd) stars become higher 
by $\approx 0.1$~mag, if we use the same PLC for the LMC as in the 
present paper. Nevertheless, these revised distance moduli computed 
from the RRd populations of M15, M68 and IC4499 are still in the range 
of $18.54$--$18.61$, which range is consistent with the value based 
on the present BW analysis. 
  
We mention the recent effort of Cacciari et al. (2000) in revisiting 
the BW analysis of RR~Lyrae stars with updated physics. From their 
analysis of two stars they concluded that the derived luminosities 
were in agreement with the earlier results (meaning the continuance 
of the old discrepancy, however, see C03 for a different conclusion 
from the same work).

%%%%%%%%%%%%%%%%%%%%%
%  SECTION 5
%%%%%%%%%%%%%%%%%%%%%

\section{Conclusions}
We have shown that by using exactly the same method and Baade-Wesselink 
parameters (e.g., velocity projection factor), fundamental mode RR~Lyrae 
and Cepheid variables yield the same distance modulus for the Large 
Magellanic Cloud. This is in variance with the earlier common belief 
that the two groups of stars lead to significantly different distance 
estimates for the LMC. The higher luminosity obtained for the RR~Lyrae 
stars in the present study is attributed to: (a) the more accurate 
consideration of phase-dependent bolometric correction, (b) the more 
precise computation of the temporal gravity, (c) the better numerical 
treatment of both the grids of stellar atmosphere models and the 
observed light and velocity curves. The good agreement between the 
Cepheid luminosities obtained in this and in the earlier works 
suggests that the above conditions have already been met in those 
studies.

Although clarification of important theoretical problems (such as 
the accurate conversion of the projected velocity to the photospheric 
value, or the application of dynamical atmosphere models rather than 
static ones) are still needed, and better agreement between various 
photometric data on LMC is still demanded, we think that the present 
result is encouraging and indicates the possibility of making progress 
either by increasing the sample size by more precise data, or by 
building dynamical atmosphere models.

\section*{Acknowledgments}
We are grateful to Fiorella Castelli for consulting on the stellar 
atmosphere models and to Andrzej Udalski for clarifying notation in 
the PLC relation. Thanks are also due to L\'aszl\'o Szabados for 
useful discussions on Cepheids, to Sza\-bolcs Barcza for careful reading 
of the manuscript and to the anonymous referee for the constructive 
comments. The support of {\sc otka} grant {\sc t$-$038437} is acknowledged.

%\bsp

\label{lastpage}

\end{document}